\documentclass[aps,twocolumn,floats,prd,nofootinbib,,superscriptaddress,letterpaper]{revtex4-1} 

\usepackage{physics}
\usepackage{enumerate}
\usepackage{empheq}
\usepackage{booktabs}
\usepackage{amsmath,amssymb,amsbsy,amstext, amsthm}
\usepackage{hyperref}
\usepackage{graphicx}
\usepackage{amsfonts}
\usepackage{color}
\usepackage{framed}
\usepackage{wasysym}

\usepackage{colortbl}
\definecolor{summersky}{cmyk}{0.71,0.33,0,0.5}
\definecolor{flamingo}{cmyk}{0,0.51,0.71,0.5}
\definecolor{rp}{cmyk}{0.2, 1, 0.6, 0}
\definecolor{pacificblue}{cmyk}{0.95,0.3,0, 0.5}
\definecolor{gray60}{cmyk}{0.4,0.4,0,0.8}
\definecolor{taiga}{cmyk}{0.68,0,0.56,0.49}

\hypersetup{
    pdftoolbar=true,        
    pdfmenubar=true,        
    pdffitwindow=true,     
    pdfstartview={FitH},    
    pdfnewwindow=true,      
    colorlinks=true,       
    linkcolor=pacificblue,          
    citecolor=flamingo,        
    filecolor=magenta,      
    urlcolor=pacificblue           
}

\newcommand{\be}{\begin{align} }
\newcommand{\ee}{\end{align} }
\newcommand{\bs}{\begin{split} }
\newcommand{\es}{\end{split} }

\newcommand{\rd}{\mathrm{d} }

\newcommand{\nn}{\nonumber\\}
\newcommand{\bfk}{{\bf k}}
\newcommand{\bfq}{{\bf q}}

\newcommand{\Mpl}{M_{\text{Pl}}}

\newcommand{\e}{\epsilon}
\newcommand{\cs}{c_{s}^{2}}

\begin{document}

\title{Soft Theorems For Shift-Symmetric Cosmologies}

\author{Bernardo Finelli}
\author{Garrett Goon}\affiliation{Institute for Theoretical Physics and Center for Extreme Matter and Emergent Phenomena,
Utrecht University, Leuvenlaan 4, 3584 CE Utrecht, The Netherlands}
\author{Enrico Pajer}
\affiliation{Institute for Theoretical Physics and Center for Extreme Matter and Emergent Phenomena,
Utrecht University, Leuvenlaan 4, 3584 CE Utrecht, The Netherlands}
\author{Luca Santoni}
\affiliation{Institute for Theoretical Physics and Center for Extreme Matter and Emergent Phenomena,
Utrecht University, Leuvenlaan 4, 3584 CE Utrecht, The Netherlands}

\begin{abstract}
We derive soft theorems for single-clock cosmologies that enjoy a shift symmetry. These so-called consistency conditions arise from a combination of a large diffeomorphism and the internal shift symmetry and fix the squeezed limit of all correlators with a soft scalar mode. As an application, we show that our results reproduce the squeezed bispectrum for Ultra-slow-roll inflation, a particular shift-symmetric, non-attractor model which is known to violate Maldacena's consistency relation. Similar results have been previously obtained by Mooij and Palma using background-wave methods. Our results shed new light on the infrared structure of single-clock cosmological spacetimes.

\end{abstract}

\maketitle

 
\section{Introduction}\label{ssec:}

Model independent results based on symmetries are precious because they allow us to robustly discriminate large classes of a microscopic realization of inflation. A prominent example is Maldacena's consistency relations \cite{Maldacena:2002vr} for soft scalars and gravitons. While these soft theorems are valid in very general single-clock models of inflation \cite{Creminelli:2004yq}, there exist non-attractor models in which they are violated for some extended interval of time (and therefore scales) \cite{Namjoo:2012aa,Chen:2013aj,Martin:2012pe}. It is important to note that the symmetries underlying soft theorems are the residual asymptotic symmetries, also known as Weinberg's adiabatic modes (WAMs) \cite{Weinberg:2003sw}, which are common to \textit{any} Friedmann-Lemaitre-Robertson-Walker (FLRW) spacetime \cite{Hinterbichler:2012nm} and are not specific to inflation (quasi-de Sitter expansion). Further, bona fide microscopic realizations of inflation very generally come with an additional internal symmetry, typically a \textit{shift symmetry}, to facilitate an extended period of accelerated expansion. In this paper, we show that, in the presence of such a shift symmetry, the dynamics of soft modes in non-attractor inflation is constrained in a way similar to standard attractor models. We start in Sec.$\!$ \ref{sec:ShiftAdiabaticModes} by deriving shift-symmetric adiabatic modes as extensions of WAMs . Then, in Sec.$\!$ \ref{sec:OPE}, we introduce the Operator Product Expansion, fix its leading coefficients by symmetries and use it to derive soft theorems. In Sec.$\!$ \ref{sec:Examples}, the theorems are shown to reproduce the known soft limit for Ultra-slow-roll inflation. We conclude with an outlook in Sec.$\!$ \ref{sec:outlook}.

 
\section{Shifty adiabatic modes}\label{sec:ShiftAdiabaticModes}

 In this section, we discuss a new symmetry of the action for gravity plus a scalar field $\Phi$ around any FLRW background which arises when the underlying UV-theory obeys
\begin{align}
 \Phi\to\Phi+c\label{ShiftSymmetry0}\,,
 \end{align} with $c=$constant. $  P(X) $ superfluids and higher derivative models such as G-inflation \cite{Deffayet:2010qz,Kobayashi:2010cm} fall in this class:
 \begin{align}
 \mathcal{L}=\frac{\Mpl^{2}}{2}R+P(X)+G(X)\square\phi \, , \ X\equiv-\frac{1}{2}(\nabla\Phi)^{2}\ ,\label{PandGofXTheories}
 \end{align}
 in mostly plus signature.

 The new symmetry for scalar gravitational perturbations comes from a combination of the internal shift symmetry and a large gauge transformation, which together form a residual symmetry of the gauge-fixed action. The solutions generated by these transformations are named \textit{shift-symmetric adiabatic modes} (SAMs). These exist in addition to the standard Weinberg adiabatic modes \cite{Weinberg:2003sw}. SAMs are interesting both classically and quantum-mechanically. At the classical level, SAMs represent solutions which exist for very general systems (although they might not be the solution chosen by the system). At the quantum level, SAMs generate soft theorems, which are derived in the next section.

SAMs are derived by following the standard protocol \cite{Weinberg:2003sw}:
\begin{enumerate}[(i)]
\item Use all possible small gauge symmetries to fix a gauge.
\item Find a residual symmetry transformation which preserves the gauge.
\item Demand that the field profile induced by the residual transformation solves all equations of motion  at \textit{finite} momentum, $  q\neq0$.
\item Configurations generated by transformations obeying the above are the sought after SAMs.
\end{enumerate}

We work in comoving gauge, as in \cite{Maldacena:2002vr}, where $\Phi(x^{\mu})=\bar{\Phi}(t)$,  $  \delta \Phi=0 $ and the metric takes on the form  \begin{align}
\rd s^{2}&=-\left(  1+\delta N\right)^{2}\rd t^{2}+ \nonumber\\
&\quad +a^{2}e^{2\zeta}\delta_{ij}\left( \rd x^{i}+N^{i}\rd t \right)\left( \rd x^{j}+N^{j}\rd t \right)\,,\label{ScalarMetric}
\end{align}
where, concentrating only on scalar modes $  N^{i}\equiv \partial_{i}\psi $. Spatial indices are raised and lowered with $\delta_{ij}$.

 Under a gauge transformation $  x^{\mu}\rightarrow  x^{\mu}+\xi^{\mu} $ one finds
\begin{align}
 \Delta\delta\Phi(x)=\pounds_{\xi}\Phi(x)=\xi^{0}\dot{\bar{\Phi}}\, ,\label{PhiShiftUnderDiff}
\end{align}
where $\pounds_{\xi}$ is the Lie derivative along $\xi^{\mu}$.
Preserving $\Phi=\bar{\Phi}(t)$ then generically requires $  \xi^{0}=0 $ \cite{HinterbiCHLer:2013dpa}. However, when \eqref{ShiftSymmetry0} holds, we can instead tolerate a \textit{constant} shift in $  \Phi $, as a diagonal combination of a diffeomorphism and an internal shift can produce $\Delta\Phi=0$. Thus, we allow $  \xi^{0}=c/\dot{\bar{\Phi}} $, for some infinitesimal constant $  c $.

On its own, $  \xi^{0}=c/\dot{\bar{\Phi}} $ does not generate an adiabatic mode, as the corresponding $\zeta$ profile is not a solution of $\zeta$'s equation of motion (EOM). 
  A diffeomorphism causes the metric \eqref{ScalarMetric} to change as $g_{\mu\nu}(x)\to g_{\mu\nu}'(x)$, where $g_{\mu\nu}'(x)$ takes on the form \eqref{ScalarMetric} but with $\delta N(x)\to \delta N(x)+\Delta\delta N(x)$ and similar expressions for the other perturbations, where the shifts are defined by
\begin{align}
 g'_{\mu\nu}(x)&\equiv g_{\mu\nu}(x)+ \pounds_{\xi}g_{\mu\nu}(x)\, .\label{MetricLieDerivative}
 \end{align}
  The comoving curvature perturbation corresponding to $\xi^{0}=c/\dot{\bar{\Phi}}$ is then $  \Delta \zeta=cH/\dot{\bar{\Phi}} $.  However, it is straightforward to verify that this profile is not a solution of the well-known $\zeta$ action.  This is a symptom of the fact that the profile does not solve the EOM at finite momentum and hence fails the third step in Weinberg's protocol.

This issue can be rectified by supplementing the time shift $\xi^{0}$ by a spatial diffeomorphism $  \xi^{i} $.  The required $\xi^{i}$ turns out to be an isotropic, time-dependent spatial rescaling, $ \xi^{i}=\lambda(t)x^{i}  $.

Only two components of the Einstein equations vanish trivially at $  q=0 $, namely the momentum constraint (the equation of motion for $  N^{i} $) and the $  ij $ Einstein equations,
\begin{align}
 \partial_{i}\left( \Theta \, \delta N-\dot \zeta \right)&=0\,,\label{MomentumConstraint}\\
 \partial_{i}\partial_{j}\left(  a^{2}\dot\psi+3a^{2}H\psi+\delta N+\zeta\right)&=0\label{EinsteinijEqns}\,,
\end{align}
to linear order\footnote{Notice that \cite{HinterbiCHLer:2013dpa} and \cite{Mirbabayi:2014zpa} consider the Hamiltonian constraint, but that is superfluous since it does not vanish at $  q=0 $. Also, they do not check the $  ij $ equation, which leads to the wrong time dependence for $  \xi^{i} $ in \cite{HinterbiCHLer:2013dpa}. \cite{Mirbabayi:2014zpa} Noticed this issue and corrected it by solving the equations of motion for $  \zeta  $, but that is not needed in our derivation (see \cite{toappear}).}. The form of the function $\Theta$ in \eqref{MomentumConstraint} depends
on the details of the theory.  For example, \eqref{PandGofXTheories} corresponds to $\Theta=H+\dot{\bar{\Phi}}^{3}\partial_{X}G/2\Mpl^{2}$.
These equations are non-trivially satisfied by
\begin{align}
\delta N&=\frac{1}{\Theta}\dot \zeta\,,\label{deltaNSolution}\\
\psi&=\frac{F({\bf{x}})}{a^{3}}-\frac{1}{a^{3}}\int^{t} \rd t' a(t')\left(  \delta N(t')+\zeta(t')\right)\,,\label{psiSolution}
\end{align}
for arbitrary $  F({\mathbf{x}}) $. Though the spatially constant part of $  \psi $ is immaterial at $  q=0 $, it is required for correctly extracting the physical $  q\neq 0 $ mode (see \cite{toappear}). 

We now consider the perturbations generated by $  \xi^{\mu} $ acting on an unperturbed FLRW:
\begin{align}
\Delta \delta N&=\dot\xi^{0}\, , \ \Delta \zeta=H\xi^{0}+\frac{1}{3}\partial_{i}\xi^{i}\, , \ \Delta \partial_{i} \psi=\dot{\xi}^{i}-a^{-2}\partial_{i}\xi^{0}\, ,\label{DeltaNDeltazetaPsiInduced}
\end{align}
at lowest order in fields. Substituting $\xi^{\mu}=\{ c/\dot{\bar{\Phi}},c\,\lambda(t)x^{i} \}$ into \eqref{DeltaNDeltazetaPsiInduced} to find the potential adiabatic modes and then substituting these modes into the first constraint \eqref{deltaNSolution}, it is found that the large diffeomorphisms that can be extended to physical SAMs must have
\begin{align}
\lambda(t)&=C_{1}-\int^{t} \rd t' \,\left (\frac{\dot H}{\dot{\bar{ \Phi}}} +\left (\Theta-H\right )\frac{\ddot{\bar{\Phi}}}{\dot{\bar{\Phi}}^{2}}\right )\ .\label{lambdaSolution}
\end{align}   In \eqref{lambdaSolution} $  C_{1} $ is an integration constant and, further, can be recognized as the usual, leading scalar WAM in comoving gauge.   We set $C_{1}\to 0$ in the remainder of this paper.

In summary, the shift-symmetric adiabatic modes are induced by the preceding diffeomorphism and generate the comoving curvature perturbation SAM mode \eqref{DeltaNDeltazetaPsiInduced}
\begin{align}
\Delta \zeta
&=c \left( \frac{H}{\dot{\bar{\Phi}}}+\lambda(t)\right) \ \quad \text{(SAM solution)}\, , \label{zetaSAMMode}
\end{align}
with $\lambda(t)$ as in \eqref{lambdaSolution}.
The adiabatic profiles $\Delta \delta N$ and $\Delta \psi$ are similarly determined from \eqref{DeltaNDeltazetaPsiInduced} and \eqref{psiSolution}, the latter of which is needed for finding the homogeneous part of $\Delta \psi$. This mode can further be shown to exist for a generic perfect fluid \cite{toappear}.

The addition of the second term in \eqref{zetaSAMMode} ensures that this profile solves the $\zeta$ equations of motion in the soft $q\to 0$ limit.  We have checked this explicitly to lowest order for a pure $  P(X) $ theory: the non-linear shift in $  \zeta $ must solve the linear equations of motion at finite momentum\footnote{By finite momentum we are referring to the fact that to derive these equations of motion one must use the finite momentum solutions for the $  \delta N $ and $  N^{i} $ constraint equations as in, e.g., \cite{Maldacena:2002vr}.}
\begin{align}\label{FiniteMomentumEOM}
\partial_{t}\left(  a^{3}\e c_{s}^{-2} \dot\zeta\right)=\mathcal{O}(q^{2})\,.
\end{align}
Upon repeated use of the background equations of motion, it is straightforward to verify that \eqref{zetaSAMMode} is indeed a solution. 

Finally, the SAM also represents an exact, non-linear symmetry of the gauge-fixed action which determines the Ward identities we derive in the following section.  The form of the symmetry simply follows from keeping the field dependent terms in \eqref{MetricLieDerivative} \cite{Hinterbichler:2012nm}. For $\zeta$ the result is:
\begin{align}
\Delta \zeta&=H\xi^{0}+\frac{1}{2}\vec{\partial}\xi^{0}\cdot\vec{\partial}\psi+\frac{1}{3}\partial_{i}\xi^{i}+\xi^{\mu}\partial_{\mu}\zeta\,, \label{FirstOrderDiffOfZeta}
\end{align}
with $\xi^{\mu}$ as given before.

 
\section{Soft Theorems}\label{sec:OPE}

We now use the shift symmetry and the dilation symmetry to fix the leading terms in the Operator Product Expansion (OPE).  Following \cite{Assassi:2012et,Kehagias:2012pd}, we assume the existence of a momentum space OPE whose leading terms are
\begin{align}
\zeta_{\bfk-\frac{1}{2}\bfq}\zeta_{-\bfk-\frac{1}{2}\bfq}\xrightarrow{\bfq\to 0}& \, P(k)(2\pi)^{3}\delta^{3}(\bfq)+f(k)\zeta_{-\bfq}\nn
&\quad+g(k)\dot{\zeta}_{-\bfq}+\mathcal{O}(q\zeta,\zeta^{2})\label{GeneralLinearOPE}\,,
\end{align} 
where $\zeta_{\bfk}=\zeta(\bfk,t)$, $\bfk\equiv \vec{k}$, the $\mathcal{O}(\zeta^{0})$ term is fixed by the power spectrum and $f(k)$ and $g(k)$ are two unknown functions of momentum which will be fixed by symmetries. In slow-roll inflation, it was proven \cite{Assassi:2012et} that on superhorizon scales $\dot{\zeta}$ can be written purely as a function of $\zeta$ as an operator equation and hence the second line in \eqref{GeneralLinearOPE} can be ignored in that context.  However, the non-conservation of $\zeta$ on superhorizon scales in non-attractor models of inflation violates the assumptions crucial to this replacement (as noted in \cite{Assassi:2012et}) and hence we retain $\dot{\zeta}$ as an independent operator in the OPE.

We can fix $f(k)$ and $g(k)$ using the dilation and shift symmetries, respectively.  The dilation symmetry induces 
\begin{align}
\zeta_{\bfk}\to \zeta_{\bfk}+\lambda (2\pi)^{3}\delta^{3}(\bfk)-\lambda\left (3+\bfk\cdot\partial_{\bfk}\right )\zeta_{\bfk}\label{DilationSymmetry}
\end{align}
for infinitesimal, constant $\lambda$, while the shift symmetry generates
\begin{align}
\zeta_{\bfk}\to \zeta_{\bfk}&+c \, \lambda(t)\left ((2\pi)^{3}\delta^{3}(\bfk)-\left (3+\bfk\cdot\partial_{\bfk}\right )\zeta_{\bfk}\right )  \nn
&\quad +\frac{c}{\dot{\bar{\Phi}}}\left (H(2\pi)^{3}\delta^{3}(\bfk)+\dot{\zeta}_{\bfk}\right )\,,\label{ShiftSymmetry}
\end{align}
where we used \eqref{FirstOrderDiffOfZeta} and with $\lambda(t)$ as in \eqref{lambdaSolution}. The functions $f(k)$ and $g(k)$ are then calculated by taking the expectation value of the commutator of $\zeta_{\bfk_{1}}\zeta_{\bfk_{2}}$ with the charges $Q_{i}$ which generate the symmetries \eqref{DilationSymmetry} and \eqref{ShiftSymmetry} and then taking the OPE limit where \eqref{GeneralLinearOPE} can be used.  For instance, in $\langle \left [iQ_{D},\zeta_{\bfk_{1}}\zeta_{\bfk_{2}}\right ]\rangle$ with $Q_{D}$ the dilatation charge operator, the only non-vanishing contribution comes from the $\mathcal{O}(\zeta)$ term in \eqref{DilationSymmetry}:
\begin{align}
\langle \left [iQ_{D},\zeta_{\bfk_{1}}\zeta_{\bfk_{2}}\right ]\rangle&=(1-n_{s})P(k_{1})(2\pi)^{3}\delta^{3}(\bfk_{1}+\bfk_{2})\ .\label{DilOPE1}
\end{align}
In the OPE limit \eqref{GeneralLinearOPE}, this can be equivalently calculated as
\begin{align}
 \langle \left [iQ_{D},\zeta_{\bfk_{1}}\zeta_{\bfk_{2}}\right ]\rangle\xrightarrow{\rm OPE}&\langle \left [iQ_{D},f(k)\zeta_{-\bfq}+g(k)\dot{\zeta}_{-\bfq}\right ]\rangle\nn
 =&\langle f(k)\Delta\zeta_{-\bfq}+g(k)\partial_{t}\Delta\zeta_{-\bfq}\rangle\nn
 =&f(k)(2\pi)^{3}\delta^{3}(\bfq)\label{DilOPE2}
\end{align}
where only the $\mathcal{O}(\zeta^{0})$ part of \eqref{DilationSymmetry} contributed non-trivially and higher order terms in the OPE were ignored.  Matching \eqref{DilOPE1} and \eqref{DilOPE2} fixes $f(k)$:
 \begin{align}
    f(k)=(1-n_{s})P(k) \ .\label{fSolution}
   \end{align} 
   By repeating the same procedure with $Q_{S}$, the generator of the shift symmetry \eqref{ShiftSymmetry}, one fixes also $g(k)$ 
   \begin{align}
   g(k)&=\frac{1}{\Theta}\frac{\dot{\bar{\Phi}}}{\ddot{\bar{\Phi}}}\left[(1-n_{s})P(k)H-\dot{P}(k) \right]\ .\label{gSolution}
   \end{align}
 As we will see, this OPE fixes the squeezed limit of the bispectrum. To determine the soft limit of higher-point correlators, we need to use the multi-field OPE \cite{Weinberg:1995mt} 
   \begin{align}
   \prod_{a=1}^{n}\zeta_{\bfk_{a}-\bfq/n}&\xrightarrow{\bfq\to 0}\sum_{\mathcal{O}}f_{\mathcal{O}}(k_{a})\mathcal{O}(-\bfq)\,,
   \end{align}
   where $\sum_{a}\bfk_{a}=0$. The lowest order term in the sum is fixed by the $n$-point correlator:
   \begin{align}
   \sum_{\mathcal{O}}f_{\mathcal{O}}(k_{a})\mathcal{O}(-\bfq)&\supset B_{n}(\bfk_{a},t)(2\pi)^{3}\delta^{3}(\bfq)\ ,
   \end{align}
   where $\langle \zeta_{\bfk_{1}}\ldots\zeta_{\bfk_{n}}\rangle\equiv B_{n}(\bfk_{a},t)(2\pi)^{3}\delta^{3}(\sum_{a}\bfk_{a})$.  Through arguments analogous to those used in determining \eqref{fSolution} and \eqref{gSolution}, we can use the dilation and shift symmetries to fix the linear terms in the generalized OPE:
   \begin{align}
   \sum_{\mathcal{O}}f_{\mathcal{O}}(k_{a})\mathcal{O}(-\bfq)&\supset f_{n}(\bfk_{a},t)\zeta_{-\bfq}+g_{n}(\bfk_{a})\dot{\zeta}_{-\bfq}\,,\nn
   f_{n}(\bfk_{a},t)&=-\mathcal{D}^{(n)}B_{n}(\bfk_{a},t)\,,\nn
   g_{n}(\bfk_{a},t)&=\frac{\dot{\bar{\Phi}}}{\ddot{\bar{\Phi}}\Theta}\left[-H\mathcal{D}^{(n)}B_{n}(\bfk_{a},t)-\dot{B}_{n}(\bfk_{a},t)\right]\,,\nn
   \mathcal{D}^{(n)}&\equiv \left [3(n-1)+\sum_{a=1}^{n}\bfk_{a}\cdot\partial_{\bfk_{a}}\right ]\ .\label{GeneralizedOPESolution}
   \end{align}

From \eqref{GeneralizedOPESolution}, soft theorems of correlators follow immediately by applying the OPE to the correlator:
\begin{align}
&\quad\lim _{\bfq\to 0}\frac{1}{2}\langle \{\zeta_{\bfq},\prod_{a=1}^{n}\zeta_{\bfk_{a}-\bfq/n}\}\rangle'\nn
&\simeq g_{n}(\bfk_{a},t) \frac{1}{2}\left (\langle \dot{\zeta}_{\bfq}\zeta_{-\bfq}+\zeta_{\bfq}\dot\zeta_{-\bfq}\rangle'\right )
 +f_{n}(\bfk_{a},t) \langle \zeta_{\bfq}\zeta_{-\bfq}\rangle'\nn
&=-\frac{\dot{\bar{\Phi}}\dot{P}(q)}{2\ddot{\bar{\Phi}}\Theta}\left [H\mathcal{D}^{(n)}B_{n}(\bfk_{a},t)+\dot{B}_{n}(\bfk_{a},t)\right ]  \nn
&\quad-P(q)\mathcal{D}^{(n)}B_{n}(\bfk_{a},t) + \mathcal{O}(q)\,, \label{SqueezedNPointFn}
\end{align}
where primes on correlators indicate that the $(2\pi)$ factors and momentum-conserving delta function have been removed.  The anti-commutator is taken in order to isolate the real part of $\langle \dot{\zeta}_{-\bfq}\zeta_{\bfq}\rangle'$. The soft theorem in \eqref{SqueezedNPointFn} is our main result. 

As a limiting case, the squeezed limit of the bispectrum reads
\begin{align}
&\quad\lim _{\bfq\to 0}\langle \zeta_{\bfq}\zeta_{\bfk-\frac{1}{2}\bfq}\zeta_{-\bfk-\frac{1}{2}\bfq}\rangle'\nn
&=-\frac{\dot{\bar{\Phi}}\dot{P}(q)}{2\ddot{\bar{\Phi}}\Theta} \left [(n_{s}-1) H P(k)+\dot{P}(k)\right ]\nn
&\quad+ (1-n_{s})P(k)P(q)\ .\label{SqueezedBispectrum}
\end{align}
For attractor inflation $  \dot P\simeq 0 $ and so the second line dominates\footnote{In standard slow-roll inflation $  \dot P/ P\propto q^{2} $ and so the first line of \eqref{SqueezedBispectrum} is of the same order as the $  \mathcal{O}(q^{2}) $ corrections that we neglected in \eqref{GeneralLinearOPE}.} and reproduces Maldacena's result. For non-attractor inflation $  \dot P \simeq \mathcal{O}(1)HP $ and so the first line will dominate when the prefactor is also $\mathcal{O}(1)$. Note, that so, far we have not used the background equations of motion nor have we expanded in any slow-roll parameters.

 
\section{Ultra-Slow-Roll Inflation}\label{sec:Examples}

In this section, we initially restrict our attention to the shift-symmetric $P(X)$ theories and then further specialize to the case of Ultra-slow-roll inflation (USR) \cite{Kinney:2005vj}.

 $P(X)$ theories are described by \eqref{PandGofXTheories} with $G(X)= 0$.  The background equations of motion can be combined to give the following relation among Hubble slow-roll parameters  $  \e\equiv -\dot H/H^{2} $, $  \eta\equiv \dot\e/(\e H) $ and the speed of sound $  c_{s}^{2}\equiv P_{,X}/(P_{,X}+2XP_{,XX}) $:
\begin{align}
\e \left( 3+\frac{\eta-2\e}{1+\cs} \right)=0\,.
\end{align}
This demonstrates that  \textit{standard slow-roll inflation is impossible for pure $  P(X) $ theories}.  Here, ``slow-roll" refers to the standard conditions $0<\e\ll 1$ and $|\eta|\ll 1$.  Despite this fact, it is still possible to generate a nearly scale invariant power spectrum, for instance as in the ghost condensate \cite{ArkaniHamed:2003uy} (corresponding to $\epsilon=0$) or as in USR, which we review next.

We now verify the prediction for the squeezed bispectrum \eqref{SqueezedBispectrum} in the case of USR inflation, which is the limit where $P(X)=X+{\rm constant}$, for which $c_s=1$.  USR admits a background for which $\epsilon\ll 1$, $\eta\simeq -6$ and whose two- and three-point functions are given by
\begin{align}
 \langle\zeta_{\bfk}\zeta_{-\bfk}\rangle'&=\frac{H^{2}}{4M_{p}^{2}\epsilon k^{3}}\,,\nn
 \langle \zeta_{\bfk_{1}}\zeta_{\bfk_{2}}\zeta_{\bfk_{3}}\rangle&=\frac{3H^{4}}{16M_{p}^{4}\epsilon^{2}}\frac{\sum_{i}k_{i}^{3}}{\prod_{i}k_{i}^{3}}=3\sum_{j>i}P(k_{i})P(k_{j})\ .\label{USRCorrelators}
 \end{align}
 As indicated, the non-Gaussianity is completely local, even away from the squeezed limit. In the squeezed limit, the bispectrum becomes
 \begin{align}
  \lim _{\bfq\to 0}\langle \zeta_{\bfq}\zeta_{\bfk}\zeta_{-\bfk}\rangle'&=6P(k)P(q)\,,
  \end{align} which differs from the naive Maldacena prediction by a sign \cite{Namjoo:2012aa}. Note that while the power spectrum takes on the standard slow-roll form, it is now strongly time dependent: $\epsilon\propto a^{-6}$ and so $P\propto \epsilon^{-1}\propto a^{6}$. This is equivalent to the fact that $\zeta$ does not freeze on superhorizon scales in this model.
  
   A background-wave type argument was used in \cite{Mooij:2015yka} to argue for this form of the squeezed bispectrum.  Here we use the OPE result \eqref{SqueezedBispectrum}.   We simply need to plug $\dot{P}(k)\simeq 6HP(k)$, $\Theta=H$ and the background EOM $ \ddot{\bar{\Phi}}+3H \dot{\bar{\Phi}}=0$ into  \eqref{SqueezedBispectrum}. The dilation contributions cancel and the result for the bispectrum reads simply
   \begin{equation}
 \lim _{\bfq\to 0}\langle \zeta_{\bfq}\zeta_{\bfk-\frac{1}{2}\bfq}\zeta_{-\bfk-\frac{1}{2}\bfq}\rangle' \simeq 6P(k)P(q)\, ,
   \end{equation}
   corresponding to $f_{\text{NL}}=\frac{5}{2}$, in agreement with \cite{Namjoo:2012aa}.  Similar calculations for higher-point correlators are straightforward.
   
     While both our argument and that of \cite{Mooij:2015yka} reproduce the correct USR bispectrum, our general expressions for the squeezed limit are quite different.  For instance, ignoring the terms proportional to the scalar tilt in \eqref{SqueezedBispectrum}, our result reads $\langle \zeta^{3}\rangle\sim H^{-1}\dot{P}^{2}$ while theirs is of the form $\langle \zeta^{3}\rangle\sim P\dot{P}$.   It would be interesting to test both of our methods in other models of shift-symmetric non-attractor inflation; this is left for later work.
 
\section{Discussion}\label{sec:outlook}

Adiabatic modes are heralded, as they both guide the classical evolution of cosmological perturbations and determine the Ward identities which relate different cosmological correlators. 
In this paper, we have demonstrated that a \textit{new} adiabatic mode arises when the scalar field (the clock) further enjoys an internal shift symmetry.  We have derived the form of this shift adiabatic mode and, using the Operator Product Expansion, determined the corresponding soft theorems among its $n$-point functions.  These were verified in the concrete model of Ultra-slow-roll inflation. A few comments are in order:
\begin{itemize}
\item Our general soft theorems \eqref{SqueezedNPointFn} can be tested observationally only in the well-known case of attractor inflation, $  \dot P\simeq 0 $, when they reduce to Maldacena's soft theorem and its generalizations \cite{Maldacena:2002vr,HinterbiCHLer:2013dpa}. Away from the attractor limit, our relations become model-dependent due to the appearance non-negligible time derivatives of correlators.   These derivatives are evaluated at early times and are hence unobservable, as is the model-dependent factor $ \dot{\bar{\Phi}}/\ddot{\bar{\Phi}}\Theta$ which also enters the relations in such non-attractor scenarios. Heuristically, model-dependence arises from the fact that there is only one way in which the background can be an attractor, but many different ways in which it can be non-attractor.
\item There are two pathways to extract the consequences of a given (non-linearly realized) symmetry such as the one we consider here: either i) directly derive the Ward identities for correlators or ii) build a symmetric Lagrangian and compute the correlators from it. We will pursue the latter approach in an upcoming publication \cite{Finelli:2018upr}.
\end{itemize}

\paragraph*{Note:} While finishing this paper, we became aware that \cite{Bravo:2017wyw} was also completing related work. Both our works were submitted to the arXiv on the same day, but due to an unfortunate typographical error in one of the iterations of ``shift symmetry'' our paper was put on hold and appeared only two days later.

 
\section*{Acknowledgments}\label{sec:} We would like to thank Lasha Berezhiani, Kurt Hinterbichler and Sadra Jazayeri for useful discussions. B.F., G.G.$\!$ and E.P.$\!$ are supported by the Delta-ITP consortium, a program of the Netherlands Organization for Scientific Research (NWO) that is funded by the Dutch Ministry of Education, Culture and Science (OCW). L.S.$\!$ is supported by the Netherlands Organization for Scientific Research (NWO). This work is part of the research programme VIDI with Project No.~680-47-535, which is (partly) financed by the Netherlands Organisation for Scientific Research (NWO).

\bibliographystyle{apsrev}
\bibliography{paper_USR}

\begin{thebibliography}{20}
\expandafter\ifx\csname natexlab\endcsname\relax\def\natexlab#1{#1}\fi
\expandafter\ifx\csname bibnamefont\endcsname\relax
  \def\bibnamefont#1{#1}\fi
\expandafter\ifx\csname bibfnamefont\endcsname\relax
  \def\bibfnamefont#1{#1}\fi
\expandafter\ifx\csname citenamefont\endcsname\relax
  \def\citenamefont#1{#1}\fi
\expandafter\ifx\csname url\endcsname\relax
  \def\url#1{\texttt{#1}}\fi
\expandafter\ifx\csname urlprefix\endcsname\relax\def\urlprefix{URL }\fi
\providecommand{\bibinfo}[2]{#2}
\providecommand{\eprint}[2][]{\url{#2}}

\bibitem[{\citenamefont{Maldacena}(2003)}]{Maldacena:2002vr}
\bibinfo{author}{\bibfnamefont{J.~M.} \bibnamefont{Maldacena}},
  \bibinfo{journal}{JHEP} \textbf{\bibinfo{volume}{05}}, \bibinfo{pages}{013}
  (\bibinfo{year}{2003}), \eprint{astro-ph/0210603}.

\bibitem[{\citenamefont{Creminelli and Zaldarriaga}(2004)}]{Creminelli:2004yq}
\bibinfo{author}{\bibfnamefont{P.}~\bibnamefont{Creminelli}} \bibnamefont{and}
  \bibinfo{author}{\bibfnamefont{M.}~\bibnamefont{Zaldarriaga}},
  \bibinfo{journal}{JCAP} \textbf{\bibinfo{volume}{0410}}, \bibinfo{pages}{006}
  (\bibinfo{year}{2004}), \eprint{astro-ph/0407059}.

\bibitem[{\citenamefont{Namjoo et~al.}(2013)\citenamefont{Namjoo, Firouzjahi,
  and Sasaki}}]{Namjoo:2012aa}
\bibinfo{author}{\bibfnamefont{M.~H.} \bibnamefont{Namjoo}},
  \bibinfo{author}{\bibfnamefont{H.}~\bibnamefont{Firouzjahi}},
  \bibnamefont{and} \bibinfo{author}{\bibfnamefont{M.}~\bibnamefont{Sasaki}},
  \bibinfo{journal}{Europhys. Lett.} \textbf{\bibinfo{volume}{101}},
  \bibinfo{pages}{39001} (\bibinfo{year}{2013}), \eprint{1210.3692}.

\bibitem[{\citenamefont{Chen et~al.}(2013)\citenamefont{Chen, Firouzjahi,
  Namjoo, and Sasaki}}]{Chen:2013aj}
\bibinfo{author}{\bibfnamefont{X.}~\bibnamefont{Chen}},
  \bibinfo{author}{\bibfnamefont{H.}~\bibnamefont{Firouzjahi}},
  \bibinfo{author}{\bibfnamefont{M.~H.} \bibnamefont{Namjoo}},
  \bibnamefont{and} \bibinfo{author}{\bibfnamefont{M.}~\bibnamefont{Sasaki}},
  \bibinfo{journal}{Europhys. Lett.} \textbf{\bibinfo{volume}{102}},
  \bibinfo{pages}{59001} (\bibinfo{year}{2013}), \eprint{1301.5699}.

\bibitem[{\citenamefont{Martin et~al.}(2013)\citenamefont{Martin, Motohashi,
  and Suyama}}]{Martin:2012pe}
\bibinfo{author}{\bibfnamefont{J.}~\bibnamefont{Martin}},
  \bibinfo{author}{\bibfnamefont{H.}~\bibnamefont{Motohashi}},
  \bibnamefont{and} \bibinfo{author}{\bibfnamefont{T.}~\bibnamefont{Suyama}},
  \bibinfo{journal}{Phys. Rev.} \textbf{\bibinfo{volume}{D87}},
  \bibinfo{pages}{023514} (\bibinfo{year}{2013}), \eprint{1211.0083}.

\bibitem[{\citenamefont{Weinberg}(2003)}]{Weinberg:2003sw}
\bibinfo{author}{\bibfnamefont{S.}~\bibnamefont{Weinberg}},
  \bibinfo{journal}{Phys. Rev.} \textbf{\bibinfo{volume}{D67}},
  \bibinfo{pages}{123504} (\bibinfo{year}{2003}), \eprint{astro-ph/0302326}.

\bibitem[{\citenamefont{Hinterbichler et~al.}(2012)\citenamefont{Hinterbichler,
  Hui, and Khoury}}]{Hinterbichler:2012nm}
\bibinfo{author}{\bibfnamefont{K.}~\bibnamefont{Hinterbichler}},
  \bibinfo{author}{\bibfnamefont{L.}~\bibnamefont{Hui}}, \bibnamefont{and}
  \bibinfo{author}{\bibfnamefont{J.}~\bibnamefont{Khoury}},
  \bibinfo{journal}{JCAP} \textbf{\bibinfo{volume}{1208}}, \bibinfo{pages}{017}
  (\bibinfo{year}{2012}), \eprint{1203.6351}.

\bibitem[{\citenamefont{Deffayet et~al.}(2010)\citenamefont{Deffayet, Pujolas,
  Sawicki, and Vikman}}]{Deffayet:2010qz}
\bibinfo{author}{\bibfnamefont{C.}~\bibnamefont{Deffayet}},
  \bibinfo{author}{\bibfnamefont{O.}~\bibnamefont{Pujolas}},
  \bibinfo{author}{\bibfnamefont{I.}~\bibnamefont{Sawicki}}, \bibnamefont{and}
  \bibinfo{author}{\bibfnamefont{A.}~\bibnamefont{Vikman}},
  \bibinfo{journal}{JCAP} \textbf{\bibinfo{volume}{1010}}, \bibinfo{pages}{026}
  (\bibinfo{year}{2010}), \eprint{1008.0048}.

\bibitem[{\citenamefont{Kobayashi et~al.}(2010)\citenamefont{Kobayashi,
  Yamaguchi, and Yokoyama}}]{Kobayashi:2010cm}
\bibinfo{author}{\bibfnamefont{T.}~\bibnamefont{Kobayashi}},
  \bibinfo{author}{\bibfnamefont{M.}~\bibnamefont{Yamaguchi}},
  \bibnamefont{and} \bibinfo{author}{\bibfnamefont{J.}~\bibnamefont{Yokoyama}},
  \bibinfo{journal}{Phys. Rev. Lett.} \textbf{\bibinfo{volume}{105}},
  \bibinfo{pages}{231302} (\bibinfo{year}{2010}), \eprint{1008.0603}.

\bibitem[{\citenamefont{Hinterbichler et~al.}(2014)\citenamefont{Hinterbichler,
  Hui, and Khoury}}]{HinterbiCHLer:2013dpa}
\bibinfo{author}{\bibfnamefont{K.}~\bibnamefont{Hinterbichler}},
  \bibinfo{author}{\bibfnamefont{L.}~\bibnamefont{Hui}}, \bibnamefont{and}
  \bibinfo{author}{\bibfnamefont{J.}~\bibnamefont{Khoury}},
  \bibinfo{journal}{JCAP} \textbf{\bibinfo{volume}{1401}}, \bibinfo{pages}{039}
  (\bibinfo{year}{2014}), \eprint{1304.5527}.

\bibitem[{\citenamefont{Mirbabayi and Zaldarriaga}(2015)}]{Mirbabayi:2014zpa}
\bibinfo{author}{\bibfnamefont{M.}~\bibnamefont{Mirbabayi}} \bibnamefont{and}
  \bibinfo{author}{\bibfnamefont{M.}~\bibnamefont{Zaldarriaga}},
  \bibinfo{journal}{JCAP} \textbf{\bibinfo{volume}{1503}}, \bibinfo{pages}{025}
  (\bibinfo{year}{2015}), \eprint{1409.6317}.

\bibitem[{\citenamefont{Pajer and Jazayeri}(2017)}]{toappear}
\bibinfo{author}{\bibfnamefont{E.}~\bibnamefont{Pajer}} \bibnamefont{and}
  \bibinfo{author}{\bibfnamefont{S.}~\bibnamefont{Jazayeri}}
  (\bibinfo{year}{2017}), \eprint{1710.02177}.

\bibitem[{\citenamefont{Assassi et~al.}(2013)\citenamefont{Assassi, Baumann,
  and Green}}]{Assassi:2012et}
\bibinfo{author}{\bibfnamefont{V.}~\bibnamefont{Assassi}},
  \bibinfo{author}{\bibfnamefont{D.}~\bibnamefont{Baumann}}, \bibnamefont{and}
  \bibinfo{author}{\bibfnamefont{D.}~\bibnamefont{Green}},
  \bibinfo{journal}{JHEP} \textbf{\bibinfo{volume}{02}}, \bibinfo{pages}{151}
  (\bibinfo{year}{2013}), \eprint{1210.7792}.

\bibitem[{\citenamefont{Kehagias and Riotto}(2012)}]{Kehagias:2012pd}
\bibinfo{author}{\bibfnamefont{A.}~\bibnamefont{Kehagias}} \bibnamefont{and}
  \bibinfo{author}{\bibfnamefont{A.}~\bibnamefont{Riotto}},
  \bibinfo{journal}{Nucl. Phys.} \textbf{\bibinfo{volume}{B864}},
  \bibinfo{pages}{492} (\bibinfo{year}{2012}), \eprint{1205.1523}.

\bibitem[{\citenamefont{Weinberg}(2005)}]{Weinberg:1995mt}
\bibinfo{author}{\bibfnamefont{S.}~\bibnamefont{Weinberg}},
  \emph{\bibinfo{title}{{The Quantum theory of fields. Vol. 1: Foundations}}}
  (\bibinfo{publisher}{Cambridge University Press}, \bibinfo{year}{2005}), ISBN
  \bibinfo{isbn}{9780521670531, 9780511252044}.

\bibitem[{\citenamefont{Kinney}(2005)}]{Kinney:2005vj}
\bibinfo{author}{\bibfnamefont{W.~H.} \bibnamefont{Kinney}},
  \bibinfo{journal}{Phys. Rev.} \textbf{\bibinfo{volume}{D72}},
  \bibinfo{pages}{023515} (\bibinfo{year}{2005}), \eprint{gr-qc/0503017}.

\bibitem[{\citenamefont{Arkani-Hamed et~al.}(2004)\citenamefont{Arkani-Hamed,
  Cheng, Luty, and Mukohyama}}]{ArkaniHamed:2003uy}
\bibinfo{author}{\bibfnamefont{N.}~\bibnamefont{Arkani-Hamed}},
  \bibinfo{author}{\bibfnamefont{H.-C.} \bibnamefont{Cheng}},
  \bibinfo{author}{\bibfnamefont{M.~A.} \bibnamefont{Luty}}, \bibnamefont{and}
  \bibinfo{author}{\bibfnamefont{S.}~\bibnamefont{Mukohyama}},
  \bibinfo{journal}{JHEP} \textbf{\bibinfo{volume}{05}}, \bibinfo{pages}{074}
  (\bibinfo{year}{2004}), \eprint{hep-th/0312099}.

\bibitem[{\citenamefont{Mooij and Palma}(2015)}]{Mooij:2015yka}
\bibinfo{author}{\bibfnamefont{S.}~\bibnamefont{Mooij}} \bibnamefont{and}
  \bibinfo{author}{\bibfnamefont{G.~A.} \bibnamefont{Palma}},
  \bibinfo{journal}{JCAP} \textbf{\bibinfo{volume}{1511}}, \bibinfo{pages}{025}
  (\bibinfo{year}{2015}), \eprint{1502.03458}.

\bibitem[{\citenamefont{Finelli et~al.}(2018)\citenamefont{Finelli, Goon,
  Pajer, and Santoni}}]{Finelli:2018upr}
\bibinfo{author}{\bibfnamefont{B.}~\bibnamefont{Finelli}},
  \bibinfo{author}{\bibfnamefont{G.}~\bibnamefont{Goon}},
  \bibinfo{author}{\bibfnamefont{E.}~\bibnamefont{Pajer}}, \bibnamefont{and}
  \bibinfo{author}{\bibfnamefont{L.}~\bibnamefont{Santoni}}
  (\bibinfo{year}{2018}), \eprint{1802.01580}.

\bibitem[{\citenamefont{Bravo et~al.}(2017)\citenamefont{Bravo, Mooij, Palma,
  and Pradenas}}]{Bravo:2017wyw}
\bibinfo{author}{\bibfnamefont{R.}~\bibnamefont{Bravo}},
  \bibinfo{author}{\bibfnamefont{S.}~\bibnamefont{Mooij}},
  \bibinfo{author}{\bibfnamefont{G.~A.} \bibnamefont{Palma}}, \bibnamefont{and}
  \bibinfo{author}{\bibfnamefont{B.}~\bibnamefont{Pradenas}}
  (\bibinfo{year}{2017}), \eprint{1711.02680}.

\end{thebibliography}

\end{document}